\documentclass[twocolumn,showpacs,showkeys,preprintnumbers,floatfix,letterpaper,prl]{revtex4}

\usepackage{citesort,enumerate}
\usepackage{amsmath,amssymb}  
\usepackage{bm}               
\usepackage{amscd}            
\usepackage{graphicx}         
\usepackage{hhline,multirow}  
\usepackage{dcolumn}          
\newcommand{\beq}{\begin{equation}}
\newcommand{\eeq}{\end{equation}}
\newcommand{\bea}{\begin{eqnarray}}
\newcommand{\beas}{\begin{eqnarray*}}
\newcommand{\beau}[1]{\begin{equation} \label{#1} \begin{array}{rcl}}
\newcommand{\eea}{\end{eqnarray}}
\newcommand{\eeas}{\end{eqnarray*}}
\newcommand{\eeau}{\end{array} \end{equation}}
\newcommand{\bay}{\begin{array}}
\newcommand{\eay}{\end{array}}
\newcommand{\bals}{\begin{align*}}
\newcommand{\eals}{\end{align*}}

\newcommand{\vev}[1]{\langle #1 \rangle}

\newcommand{\nbar}{{\overline n}}
\newcommand{\mbar}{{\overline m}}
\newcommand{\xbar}{{\overline x}}

\newcommand{\HH}{{\mathcal H}}

\begin{document}


\preprint{BNL-NT-07/2}

\title{Multi-nucleon correlations in Deep Inelastic Scattering
  at large Bjorken $\bm x_B$} 

\author{Alberto~Accardi$^a$, Jian-Wei~Qiu$^{a,b}$ and 
        James~P.~Vary$^a$}
\affiliation{
$^a$Department of Physics and Astronomy, Iowa State University,
Ames, IA 50011-3160, U.S.A. \\
$^b$Physics Department, Brookhaven National Laboratory, 
Upton, NY 11973-5000, U.S.A. 
}
\date{January 10, 2007}

\begin{abstract}
Using realistic nuclear many-body wave functions in the collinear
factorization framework, we evaluate nuclear Deep
Inelastic Scattering at large Bjorken $x_B$, where
correlations are needed to exceed the kinematic limits of a single
free nucleon. 
We find that recent experimental data on cross-section ratios
from the CLAS collaboration exhibit features that are not
reproduced by modern 2- and 3-nucleon correlations in the nuclear wave
function.  
New physical mechanisms are needed to explain data in
the $2 \lesssim x_B \lesssim 3$ range.
\end{abstract}



\maketitle


{\bf Introduction.}
The CLAS collaboration has recently published data on ratios 
of cross-sections for scattering a high-energy electron 
on a nucleus $A$ to $^3He$ at large $x_B<3$ \cite{Egiyan:2005hs,
Egiyan:2003vg}. 
These ratios show an interesting structure with 2 plateaus at $x_B
\gtrsim 1.5$ and $x_B \gtrsim 2.25$, of height increasing with $A$.
Earlier SLAC data on A/$^2$H ratios \cite{Frankfurt:1993sp}) and
A/$^4$He ratios \cite{Day:1988xj} 
exhibit features similar to those seen by CLAS.
The emergence of plateaus is generally interpreted as due to 2- and
3-nucleon short-range correlations in nuclei. However, the physics
underlying these correlations has remained elusive.
Short-distance multi-quark \cite{QuarkClusterModel} or multi-nucleon
clustering \cite{Frankfurt:1981mk,Frankfurt:1988nt,Egiyan:2003vg} has
been favored in several models. See
\cite{Sargsian:2002wc,Frankfurt:1988nt} for a review.   

In this paper we demonstrate the limited extent to which 
the observed plateaus can be explained in terms of realistic
inter-nucleon potentials including local short-distance repulsion 
\cite{Pieper:1992gr,CiofidegliAtti:1995qe}, which can push a
nucleon momentum to large values. The resulting 2-nucleon (NN) correlations
produce a nearly nucleus-independent shape of single nucleon momentum $l$
distributions in the $2 \lesssim l \lesssim 4$ fm$^{-1}$ range
\cite{CiofidegliAtti:1995qe,Zabolitzky}, which may explain the first
plateau. At higher momenta, 3-nucleon (NNN)
correlations become dominant \cite{Pieper:1992gr}. We investigate 
whether the NNN correlations may explain the second plateau.

Our main theoretical inputs are realistic nuclear wave functions
\cite{Pieper:1992gr,CiofidegliAtti:1995qe}, and the collinear
factorization framework to describe the scattering of the lepton on a
bound nucleon  in terms of electron-parton scatterings. 
The obtained formulas are rather general and the few approximations
are made explicit. The formalism is easily generalizable to include
other physical mechanisms.  

We provide a minimalist approach that includes
only the effect of ``Fermi motion'', or single-nucleon momentum
distributions, with exact treatment of kinematics 
at the parton, nucleon and nuclear level. It provides 
a baseline, including only nucleon degrees of
freedom and the best available nuclear wave-functions.
We find that, while the first plateau may be explained 
in our model, 
the second plateau in the $2\lesssim x_B \lesssim 3$
region is not described by inclusion of NNN
correlations. Thus, single-nucleon degrees of freedom are inadequate, 
and more exotic physical mechanisms such as those
mentioned above are needed to explain experimental data.

\begin{figure}[tb]
  \vspace*{0cm}
  \centerline{
  \includegraphics
    [width=0.6\linewidth]
    {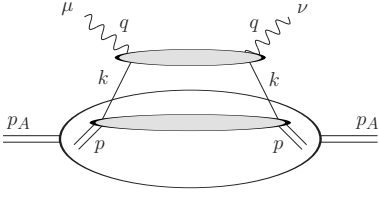}
  }
  \vspace*{0cm}
  \caption[]{
   Forward Compton amplitude for nuclear DIS.
 }
 \label{fig:forward}
\end{figure}

{\bf \indent Collinear factorization.}
Under the one-photon approximation, the cross section of nuclear 
deep inelastic scattering is determined by the
nuclear hadronic tensor,
\begin{align}
  W_A^{\mu\nu}(p_A,q) & = \frac{1}{4\pi} \int d^4z\, e^{-iq\cdot z}
    \vev{p_A|j^{\dagger\mu}(z)j^\nu(0)|p_A} \, ,
\label{wa}
\end{align}
with nuclear momentum $P_A=A\, p_A$, virtual photon momentum $q$,  
and corresponding current $j$.  
At large Bjorken $x_B=Q^2/2p_A\cdot q$ with $Q^2=-q^2$, 
the virtual photon is very localized in space-time.
In the impulse approximation, 
the virtual photon interacts with a parton of momentum $k$ belonging to
a nucleon of momentum $p$, bound in a nucleus of momentum $P_A$
and mass $M_A=A\, \mbar$, as sketched in Fig.~\ref{fig:forward}. 
We picture the nucleus as a system of $A$ bound nucleons whose 
momentum distributions (``Fermi motion'') are given by 
nuclear many-body wave functions.
If we assume the Hilbert space of the nucleus to be the direct product 
of $A$ single-nucleon Hilbert spaces, and the current operator to 
act on a single-nucleon subspace, 
\begin{align}
  W_A^{\mu\nu}(p_A,q) & \approx  \int d^4z\, \frac{1}{4\pi} \int d\mu_A\, 
    e^{-iq\cdot z} \vev{p|j^{\dagger\mu}(z)j^\nu(0)|p} \nonumber \\
  & \equiv \int d\mu_A\,  W_N^{\mu\nu}(p,q) \ ,
 \label{eq:TA_step1}
\end{align}
with $W_N^{\mu\nu}$ the hadronic tensor of a bound nucleon, and 
relativistic Fermi smearing measure, with $m^2=p^2$,
\begin{align}
  d\mu_A & = \frac{dm^2}{2\pi} \frac{d^3p}{(2\pi)^32p_0} \rho_A(p)
  |_{p_0=\sqrt{m^2+\vec p^{\,2}}}\ .
\end{align}
The single nucleon 4-momentum distribution (ND) is
\begin{align}
  \rho_A(p) & = \int \left( \prod_{i=2,A}^A 
    \frac{d^4p_i'}{(2\pi)^4} \right)
    |\phi_A(p,p_2',\ldots,p_A')|^2 \nonumber \\ 
  & \times \delta^{(4)}(p+\sum_{i=2}^A p_i' - Ap_A)
\end{align}
with $\phi_A$ the nuclear wave-function. Gauge invariance
and Eq.~\eqref{eq:TA_step1} imply $q_\mu W_A^{\mu\nu} = q_\mu
W_N^{\mu\nu} = 0$,  so
we can define the nuclear and nucleon
structure functions as follows: 
\begin{align}
  \hskip -0.05in 
  W_A^{\mu\nu}(x_B,Q^2) & = 
    - \tilde{g}^{\mu\nu} F_{1A}(x_B,Q^2) 
    + \frac{ \tilde p_A^\mu \tilde p_A^\nu}{p_A\cdot q} F_{2A}(x_B,Q^2)
    \nonumber \\
  \hskip -0.05in 
  W_N^{\mu\nu}(x_N,Q^2) & = 
    - \tilde{g}^{\mu\nu} F_{1}(x_N,Q^2) 
    + \frac{ \tilde p^\mu \tilde p^\nu}{p\cdot q} F_{2}(x_N,Q^2)
  \hskip -0.05in 
  \label{eq:nuclSFN}
\end{align}
where $x_N = Q^2/2p\cdot q$ and
$\tilde{g}^{\mu\nu} = g^{\mu\nu} - q^\mu q^\nu/q^2$,
and $\tilde a^\mu = a^\mu - (a\cdot q / q^2) q^\mu$ for any 4-vector $a$.

To evaluate the hadronic tensor or the structure functions in terms of
the QCD collinear factorization approach \cite{Collins:1989gx}, 
we need to define suitable light-cone ``+'' and
``$-$'' directions. We have 2 choices: using the $(q,p_A)$ plane
(A-frame) or using the $(q,p)$ plane (N-frame). Our goal is
to express the nuclear structure functions in terms of nucleon
structure functions. Hence, as we will discuss, 
the optimal choice is the N-frame,
\begin{align}
  p^\mu & = p^+ \nbar^\mu 
          + \frac{m^2}{2 p^+} n^\mu 
    \nonumber \\
  q^\mu & = - \xi_A\omega p^+ \nbar^\mu 
          + \frac{Q^2}{2\xi_A\omega p^+} n^\mu
   \label{eq:kinematics} \\
  p_A^\mu & = \omega p^+ \nbar^\mu 
            + \frac{\mbar_\perp^2}{2 \omega p^+} n^\mu 
            + \vec p_{A\perp}^\mu 
   \nonumber
\end{align}
where 
$\nbar = (1/\sqrt{2},\vec0_\perp,1/\sqrt{2})$, 
$n = (1/\sqrt{2},\vec0_\perp,-1/\sqrt{2})$, 
and $a^\pm = (a_0 \pm a_3)/\sqrt{2}$. In Eq.~\eqref{eq:kinematics}, 
$\omega = p_A^+ / p^+$ is the inverse nucleon 
fractional plus-momentum with respect to the nucleus, $\mbar_\perp^2 =
\mbar^2 + p_{A\perp}^2$, and
$\xi_A = 2 x_B / (1+\sqrt{1+4 x_B^2 \mbar_\perp^2 / Q^2})$
is the nuclear Nachtmann variable \cite{NachtmannVar} in the N-frame. 

Following QCD collinear factorization,  
we expand the parton momentum $k$, entering the top blob 
in Fig.~\ref{fig:forward}, around its positive light-cone component, 
$k^\mu = xp^+\nbar^\mu + O(k-xp^+\nbar)$ with 
parton momentum fraction, $x=k^+/p^+$, and factorize 
the nucleon hadronic tensor as 
\begin{align}
  W_N^{\mu\nu}(x_N,Q^2) & = \sum_f \int \frac{dx}{x} \, 
    \HH_f^{\mu\nu}(\xbar,Q^2) \,  \varphi_{f/N}(x,Q^2) 
    \nonumber \\
  & + O(1/Q^2)
 \label{eq:pQCDfact}
\end{align}
where $x_N = \xi_A \omega / (1-(\xi_A\omega)^2\, m^2/Q^2)$ and 
$\bar x = Q^2/2k\cdot q = (\xi_A\omega)/x$ in the N-frame. 
In Eq.~(\ref{eq:pQCDfact}),  
$\varphi_{f/N}$ is the leading twist PDF of a bound 
off-shell nucleon for a parton of flavor $f=g,q,\bar q$,
and $\HH_f^{\mu\nu}$ is the partonic tensor for an on-shell parton 
of momentum $k^\mu = xp^+ \nbar^\mu$ with 
all perturbative collinear divergences
along the parton momentum $k$ absorbed into the PDFs.  
With no surprise, Eq.~(\ref{eq:pQCDfact}) has the same
factorized form as that of a free nucleon \cite{Curci:1980uw}
because the factorization of short-distance partonic dynamics 
is insensitive to the details of long-distance hadron physics. 
From the parton level gauge invariance, $q_\mu \HH_f^{\mu\nu}=0$, 
we can decompose the partonic tensor as
\begin{align}
  \HH_f^{\mu\nu}(\xbar,Q^2) & = 
    - \tilde{g}^{\mu\nu}\, h_f^{1}(\xbar,Q^2) 
    + \frac{ \tilde k^\mu \tilde k^\nu}{k \cdot q}\,  
      h_f^{2}(\xbar,Q^2) \, ,
 \label{eq:partonicSFN}
\end{align}
where the scalar functions $h_f^{i}$ can be computed 
order-by-order in powers of $\alpha_s$ from the
factorized expression \eqref{eq:pQCDfact} with the nucleon state $N$
replaced by a parton state of flavor $f$, 
regardless of the nucleon state.

Combining Eqs.~\eqref{eq:TA_step1}, \eqref{eq:nuclSFN},
\eqref{eq:pQCDfact} and \eqref{eq:partonicSFN} we derive  
nuclear structure functions 
in terms of leading twist massless ($m^2=0$) nucleon 
structure functions, $F_i^{(0)}$ with $i=1,2$
\begin{widetext}
\begin{align}
  F_{2A}(x_B,Q^2) & = \frac{x_B}{1+\delta_A} \int d\mu_A
    \bigg[ 
    \frac{3(1+\delta_\omega)^2}{(1+\delta_A)(1+\delta_n)}-1 
    \bigg]
    \frac{F_2^{(0)}(\xi_A\omega,Q^2)}{2\xi_A\omega}
    \theta(1-\xi_A\omega) 
   \label{eq:F2final} \\
  F_{1A}(x_B,Q^2) & = \int d\mu_A
    \bigg\{ 
    F_1^{(0)}(\xi_A\omega,Q^2) 
    + \bigg[ 
    \frac{(1+\delta_\omega)^2}{(1+\delta_A)(1+\delta_n)}-1 
    \bigg]
    \frac{F_2^{(0)}(\xi_A\omega,Q^2)}{4\xi_A\omega}
    \bigg\}
    \theta(1-\xi_A\omega) 
  \label{eq:F1final}
\end{align}
\end{widetext}
where $\delta_A = 4x_B^2\,\mbar^2/Q^2$, $\delta_n = 4x_N^2\,
m^2/Q^2$, $\delta_\omega = 4x_Bx_N\,M_\omega^2/Q^2$,
and $M_\omega^2 = \frac{\omega}{2}
\left[m^2+\mbar_\perp^2/\omega^2\right]$.

We would like to emphasize that only in the N-frame 
we can easily generalize the free-nucleon PDF to 
the PDF of a bound off-shell nucleon because 
they both have the same operator definition.
For example, other than the state $|p\rangle$, the definition of 
quark PDF at leading order,
\begin{align}
  \varphi_q(x) = \int \frac{dz^-}{2 \pi} e^{-ixp^+z^-}
    \vev{p|\overline\psi(z^-n)\,\frac{\gamma^+}{2}\,\psi(0)|p}
\end{align}
is the same for both bound and free nucleon \cite{Collins:1981uw}. 
This would not be true in the A-frame.

\begin{figure}[tbh]
  \vspace*{0cm}
  \centerline{
  \includegraphics
    [width=0.85\linewidth,bb=18 140 590 710,clip=true]
    {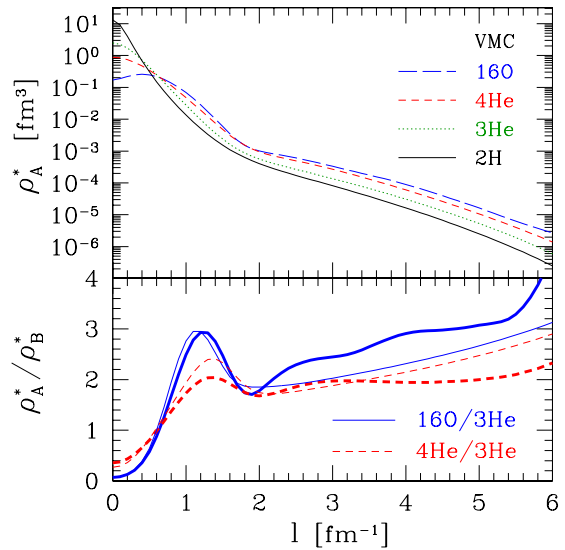}
  }
  \caption[]{
    Top: VMC nucleon momentum distributions \cite{Pieper:1992gr}. 
    Bottom: $A/^3He$ ratios
    (thick: VMC; thin: CS \cite{CiofidegliAtti:1995qe}).  
  }
 \label{fig:rho}
\end{figure}

{\bf \indent Large-$\bm x_B$ correlations.}
To apply the general result,
Eqs.~\eqref{eq:F2final}-\eqref{eq:F1final}, to describe CLAS data
we need to choose $F_i^{(0)}$ of bound nucleons and 
the measure $d\mu_A$.  First, we assume the nucleons to be on their
mass-shell with $m^2 = \mbar^2$. Thus, we may identify $F_i^{(0)}$
with the massless free nucleon structure functions. They can be
computed using PDFs from QCD global fits 
which do not already correct for the target's mass. 
Corrections for off-shell nucleons were discussed
in Ref.~\cite{Melnitchouk:1993nk} in a related, but somewhat
different, formalism. 
Next, we use non-relativistic computations of 
nucleon distribution functions
\cite{Pieper:1992gr,CiofidegliAtti:1995qe} 
(we are not aware of any relativistic model which includes NN and
NNN correlations). 
In summary we approximate 
\begin{align}
  \rho_A(p) \approx (2\pi)^4 2 p_0 \, \delta(p^2-\mbar^2) 
     \, \rho_A^{nr}(\vec{p}) \ .
\end{align}
Defining $p_A = p + l$, using the translation invariance of the
nucleon distribution, and choosing the struck nucleon rest frame,
$p=(m,\vec 0)$, we obtain
$
  d\mu_A = d^3l \rho_A^*(\vec l) \ ,
$
where $\rho_A^*(l) = \int (\prod_{i=2,A}^A d^3l_i)
    |\phi_A(l,l_2,\ldots,l_A)|^2 \delta^{(3)}(l+\sum_{i=2}^A l_i)$ 
can be identified with the nucleon distribution computed in the
nucleus rest frame. 
We consider state-of-the-art non-relativistic 
nucleon distribution functions $\rho_A^*$
obtained in a Variational Monte Carlo (VMC) computation  which 
include NN and NNN correlations \cite{Pieper:1992gr},
compared to the parametrization of $\rho_A^*$ discussed by Ciofi
degli Atti and Simula (CS) \cite{CiofidegliAtti:1995qe}, which only
considers NN correlations, see Fig.~\ref{fig:rho}. 
The relative momentum $\vec l = (\vec l_\perp,l_3)$ 
further enters Eq.~\eqref{eq:F2final} through
$\omega = (l_3 + \sqrt{l_3^2 + \mbar_\perp^2}) / m$ 
and $\mbar_\perp^2 = \mbar^2 + l_\perp^2$.
The per-nucleon nuclear deep inelastic (DIS) cross section reads
\begin{align}
  \frac{d\sigma_A}{dQ^2dx_B} & = \frac{4\pi\alpha^2}{Q^4} 
    \bigg\{
      \frac{1}{A} y^2 F_{1A}(x_B) \nonumber \\
  &   + \Big( 1-y-\frac{\mbar^2}{Q^2}x_B^2y^2 \Big)
      \frac{F_{2A}(x_B)}{x_B}
    \bigg\}
 \label{eq:xsec}
\end{align}
where $y=P_A\cdot q/ P_A\cdot p_l = Q^2/(2 E_{lab} \mbar x_B)$, $p_l$ is
the lepton initial 4-momentum and $E_{lab}$ its energy in the target
rest frame. The nuclear structure functions $F_{iA}$ are given in
Eqs.~\eqref{eq:F2final}-\eqref{eq:F1final}. 
For the following plots, we compute $F_i^{(0)}$ at leading order 
using CTEQ5L PDF \cite{Lai:1999wy}.

\begin{figure*}[tbh]
  \vspace*{0cm}
  \centerline{
  \includegraphics
    [width=0.34\linewidth]
    {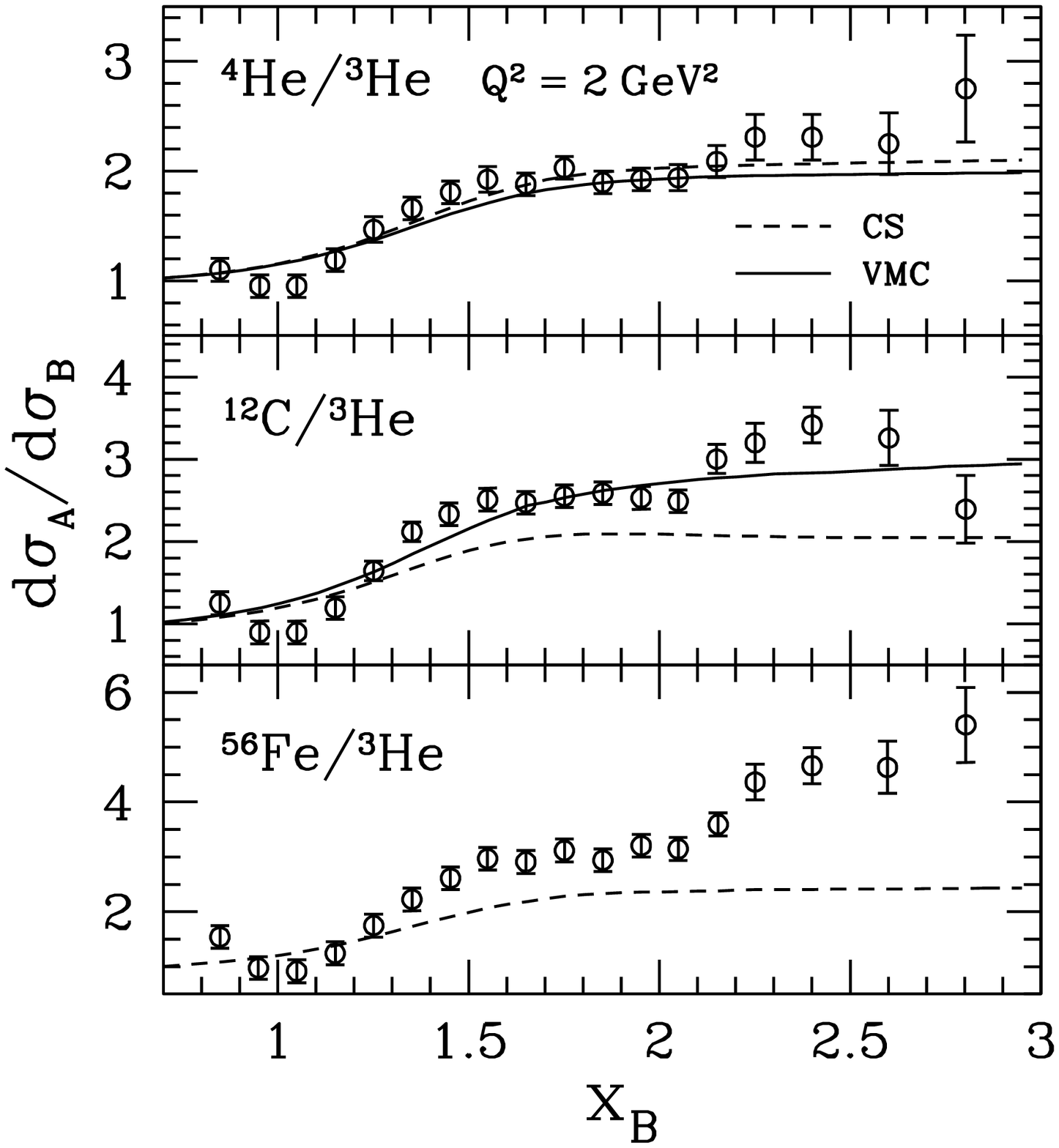}
  \includegraphics
    [width=0.34\linewidth]
    {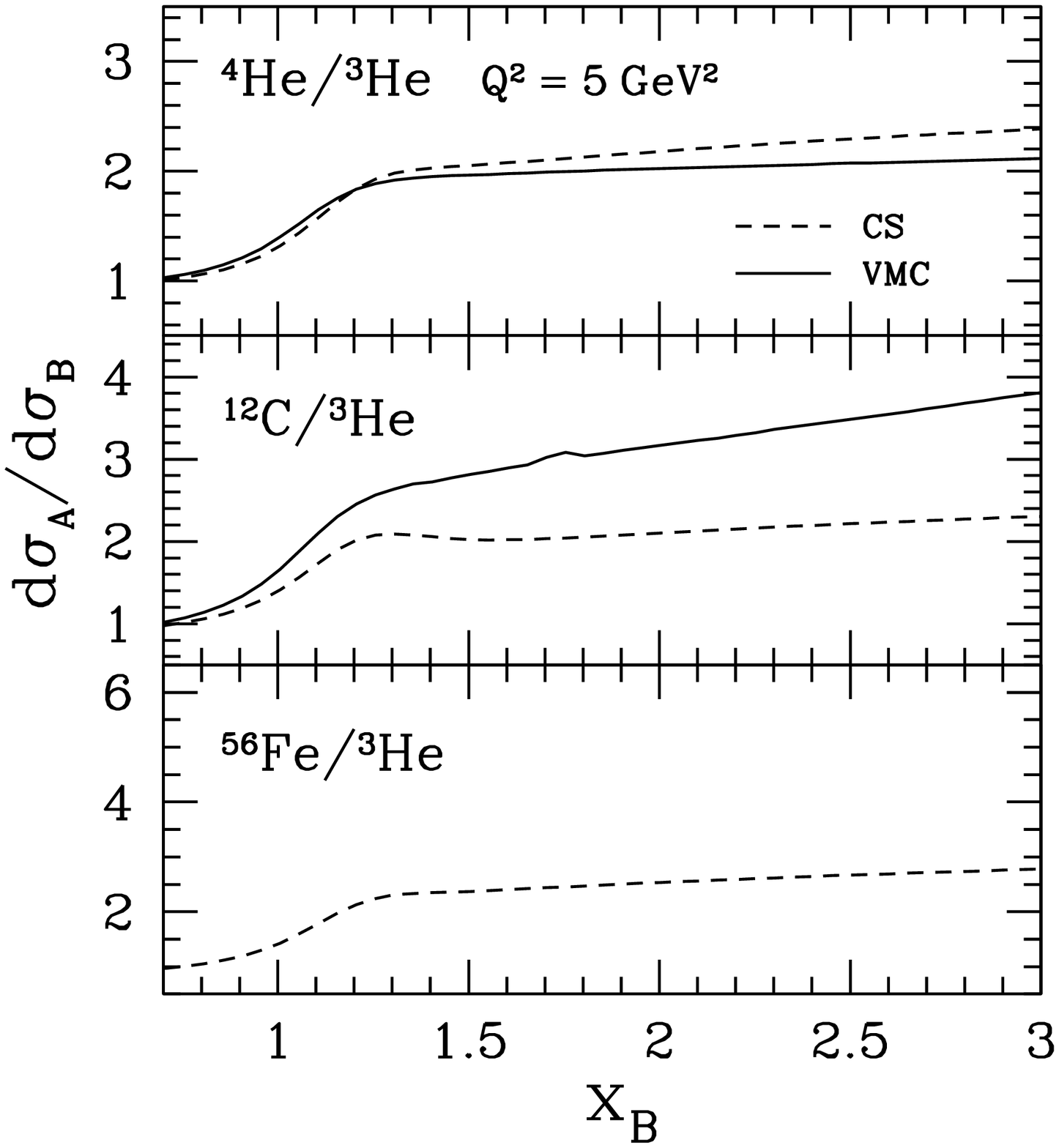}
  }
  \vspace*{0cm}
  \caption[]{
    Left: Comparison of cross-section ratios computed at $Q^2 =
    2$ GeV$^2$ and $E_{lab} = 4.5$ GeV to CLAS data taken at
    $1.6$ GeV$^2<Q^2<2.4$ GeV$^2$ \cite{Egiyan:2005hs}. Solid line
    computed with VMC nucleon distributions \cite{Pieper:1992gr}, 
    dashed line with the CS parametrization \cite{CiofidegliAtti:1995qe}.
    Right: cross-section ratios at $Q^2 = 5$ GeV$^2$ and
    $E_{lab} = 9$ GeV, accessible at the 12 GeV upgrade at JLAB.}
 \label{fig:F2ratio_JLAB}
\end{figure*}

In Fig.~\ref{fig:F2ratio_JLAB} left, we compare the $A/$${^3}He$
ratios from Eq.\eqref{eq:xsec} with CLAS data taken
at $Q^2\sim 2$ GeV$^2$ \cite{Egiyan:2005hs}. After an initial rise, the
computed ratio becomes softer starting at $x_B \gtrsim
1.5$. This is due to the onset of NN correlations which develop
a tail in the nucleon distribution at $|l|\gtrsim 2$ fm$^{-1}$, see
Fig.~\ref{fig:rho}. 
Our curves using VMC distributions (solid lines)
can describe reasonably well CLAS data for light nuclei.
However, they do not describe the onset of the second plateau, 
otherwise visible
in the ND ratios of Fig.~\ref{fig:rho} and due to the onset of NNN
correlations.  
Indeed, in the Fermi smearing of the structure
functions \eqref{eq:F2final}-\eqref{eq:F1final}, the nucleon
momentum $l$ is sampled with a rather large standard deviation of
about 1 fm$^{-1}$, which smears out such structure. 
This second plateau is then suggestive of
new mechanisms such as 6- and 9-quark clusters
\cite{QuarkClusterModel}, multi-nucleon clusters
\cite{Frankfurt:1981mk,Frankfurt:1988nt,Egiyan:2003vg},  
or dynamical short-range nucleon correlations \cite{Piasetzky:2006ai}, 
effective when nucleons approach each other at a distance smaller than
a nucleon radius.
In light nuclei such as $^3He$ or
$^4He$, all nucleons exist on the nuclear surface, which reduces the
probability of nucleon overlap. In heavier nuclei, an increasing
fraction of nucleons exist in the nuclear interior where the overlap
probability is maximized, so that we expect 
the difference between our baseline and experimental data to 
increase with $A$ at $x>2.25$.
The computation with the CS distributions (dashed lines) 
fails at $x_B \gtrsim 1.5$ 
for nuclei heavier than $^4He$, perhaps due to the lack of NNN
correlations. 

In Fig.~\ref{fig:F2ratio_JLAB} right, we show the effect of Fermi
motion on $A/{^3}He$ ratios at $Q^2=5$ GeV$^2$. The
onset of correlation effects has moved down to $x_B\sim 1$ and
the plateau's slope has increased. At
still higher values of $Q^2$ we found that the onset of correlation
effects does not move much further, but the slope of the plateau
continues to increase. 

We estimated the sensitivity of the $A/B$ ratios to different choices
of PDF, which is at most 5\%.
Parton recombination effects, which are important at small-$x_B$
\cite{Close:1989ca}, are negligible in our case.

The simple model of the nucleus we have considered can be improved in many
ways. First, we used a non-relativistic computation of $\rho_A$ to describe
the large $x_B$ region, to which nucleons of relative momentum
$l\gtrsim 2$ fm$^{-1}$ contribute: one may expect relativistic
corrections to become important and change the hard tails of the
ND. Second, the nucleon hadronic tensor can have a more
general Lorentz decomposition than in Eq.~\eqref{eq:nuclSFN}, as
discussed in \cite{Melnitchouk:1993nk} where our assumption
leading to Eq.~\eqref{eq:TA_step1} was relaxed. 
Finally, we have made no attempt to describe the EMC effect in the $0.2
\lesssim x_B \lesssim 1$ region, see
\cite{Piller:1999wx,Norton:2003cb} for a review. 
Proposed underlying mechanisms include off-shell
corrections to nucleon structure functions, nucleon binding and
removal energy corrections, and $Q^2$-rescaling. Their impact at
large-$x_B$ needs to be estimated, but we do not expect major changes
to our results since the cross-section ratios at large $x_B$ are
mainly dominated by the relative strength of the ND tails.
Quasi-elastic lepton-nucleon scattering, which causes the 
dip in experimental data at $x_B\approx 1$ can be straightforwardly
implemented, as well as contributions from the nuclear pionic cloud.
However, such improvements are not expected to make significant
contributions at large $x_B$.

Finally, at the lower $Q^2$ we considered, a few theoretical issues
arise. First, especially at $Q^2\approx 2$ GeV$^2$, one can 
question the partonic picture of DIS for computing $F_{2A}$. When
computing the ratio, however, this should not be 
an issue thanks to parton hadron duality. 
Second, one would expect higher-twist corrections to become
sizable. They would not completely cancel in the ratio, because they
are expected to be proportional to $A^{1/3}$. We will consider these
issues in a future work, but 
we expect these improvements to provide cross section ratios with 
smooth corrections that will not likely induce the second plateau
structure seen in the CLAS data.


We thank G.~Miller and S.~C.~Pieper for informative
discussions, and INT Seattle and BNL for hospitality.
We acknowledge support from DOE grant DE-FG02-87ER40371
and Contract No. DE-AC02-98CH10886.


\end{document}